\documentclass[useamsfonts,mfastrosym,usegraphicx,useepstopdf]{iau}
\usepackage{graphicx}
\usepackage{natbib}
\usepackage{wrapfig}
\usepackage{epstopdf}

\pubyear{2022}
\volume{370}
\jname{Winds of Stars and Exoplanets}
\editors{A.~A. Vidotto, L.~Fossati \& J.~Vink, eds.}

\title[Longitudinal Waves in Alfv\'{e}n wave-driven Wind] 
{Role of Longitudinal Waves in Alfv\'{e}n-wave-driven Solar/Stellar Wind}

\author[Shimizu et al.]   
{Kimihiko SHIMIZU$^1$, Munehito SHODA$^2$ \and Takeru K. SUZUKI$^{1,3}$}

\affiliation{$^1$School of Arts \& Sciences, The University of Tokyo, 3-8-1 Komaba, Meguro, Tokyo, 153-8902, Japan \\ [\affilskip]
  $^2$Department of Earth and Planetary Science, The University of Tokyo, 7-3-1, Hongo, Bunkyo, Tokyo, 113-0033, Japan \\ [\affilskip]
$^3$Department of Astronomy, The University of Tokyo, 7-3-1, Hongo, Bunkyo, Tokyo, 113-0033, Japan}

\begin{document}

\maketitle

\begin{abstract}
  We study the role the the $p$-mode-like vertical oscillation on the photosphere in driving solar winds in the framework of Alfv\'en-wave-driven winds.
  By performing one-dimensional magnetohydrodynamical numerical simulations from the photosphere to the interplanetary space, we discover that the mass-loss rate is raised up to $\approx 4$ times as the amplitude of longitudinal perturbations at the photosphere increases. 
  When the longitudinal fluctuation is added, transverse waves are generated by the mode conversion from longitudinal waves in the chromosphere, which increases Alfv\'enic Poynting flux in the corona. As a result, the coronal heating is enhanced to yield higher coronal density by the chromospheric evaporation, leading to the increase of the mass-loss rate.
Our findings clearly show the importance of the $p$-mode oscillation in the photosphere and the mode conversion in the chromosphere in determining the basic properties of the wind from the sun and solar-type stars.

\end{abstract}

\firstsection 
              
\section{Introduction}
The sun and low-mass main sequence stars posses a surface convection zone beneath the photosphere.
Various types of waves are excited and are propagating upward to the atmosphere \citep{Lighthill1952RSPSA,Stepien1988ApJ}. 
Alfv\'en waves have attracted a great attention as a reliable player in heating and driving the solar wind and stellar winds from solar-type stars because they can efficiently transport the convective kinetic energy to the corona and the wind region owing to the incompressible nature \citep{Belcher1971ApJ,Suzuki2005ApJ,Suzuki2006JGRA,Cranmer2007ApJS,Verdini2007ApJ,Matsumoto2012ApJ,Shoda2019ApJ,Sakaue2021ApJ,Matsumoto2021MNRAS,Vidotto2021LRSP}.

On the other hand, acoustic waves are not regarded to play a substantial role in heating and accelerating the solar and stellar winds because those excited by $p$-mode oscillations easily steepen to form shocks before reaching the corona \citep{Stein1972ApJ,Cranmer2007ApJS}.
Recently, however, \citet{Morton2019NatAs} reported the tight relation between the $p$-modes on the photosphere and Alfv\'enic oscillations in the corona from the comparison of their power spectra and pointed out the possibility of the generation of transverse waves by the mode conversion from longitudinal waves excited by $p$-modes \citep{Cally2011ApJ}.

Inspired by this observational finding, we studied the effect of vertical oscillations at the photosphere on Alfv\'en wave-driven solar winds by magnetohydrodynamical (MHD) simulations \citep{Shimizu2022ApJ}.
In this talk, we introduce the results of this paper and explain the importance of the longitudinal fluctuation in driving the wind from the sun and solar-type stars from a theoretical point of view.

\section{Simulation}
We set up one-dimensional (1D) super-radially open magnetic flux tubes along the $r$ direction rooted at the photosphere of a star with the solar mass, $M_{\odot}$, and radius, $R_{\odot}$.
The radial magnetic field strength is set to be $B_r = 1300$ G on the photosphere, which gives the equipartition between the gas pressure and the magnetic pressure. In the outer region after the super-radial expansion has completed, the radial field decreases with $B_r = 1.3$ G $(r/R_{\odot})^{-2}$.  
We solve compressible MHD equations with radiative cooling and thermal conduction in the flux tube from the photosphere to $\approx 100 R_{\odot}$.
We input three dimensional velocity fluctuation on the photosphere; the two transverse components, $\delta v_{\perp,1}$ and $\delta v_{\perp,2}$, generate Alfv\'en waves and the longitudinal (radial) component, $\delta v_{\parallel}$, excites acoustic waves.
In order to take into account cascading Alfv\'enic turbulence, which plays a role in the dissipation of Alfv\'en waves \citep{Goldreich1995ApJ,Matthaeus1999ApJ}, we employ phenomenological dissipation terms in the transverse components of the momentum equation and the induction equation \citep{Shoda2018ApJ}.  
See \citet{Shimizu2022ApJ} for the detailed setup of the numerical simulation.

\section{Results}
\begin{figure}[h]
\begin{center}
 \includegraphics[width=0.5\textwidth]{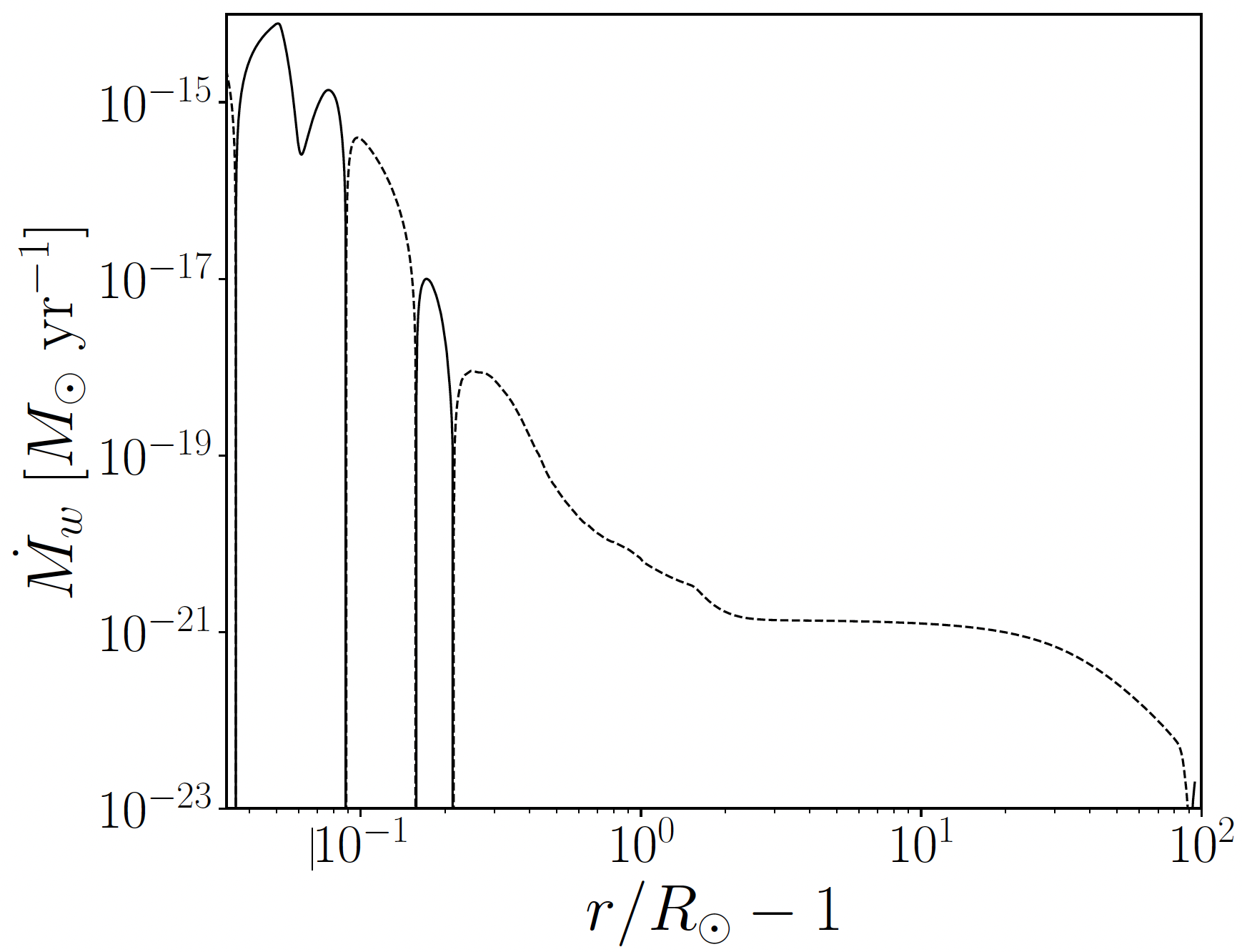} 
 \caption{Mass-loss rate (equation \ref{eq:MLR}) of the case with the only longitudinal fluctuation, $\langle \delta v_{\parallel}\rangle=0.6$ km s$^{-1}$, on the photosphere.
   The solid (dashed) line corresponds to positive (negative) $\dot{M}_w$. }
 \label{fig:onlylongitude}
\end{center}
\end{figure}
Before investigating roles of acoustic waves in Alfv\'en wave-driven winds, let us first examine whether the solar wind is driven solely by acoustic waves in the absence of Alvf\'en waves.
To do so, we run a simulation with the only longitudinal perturbation, $\langle \delta v_{\parallel}\rangle = 0.6$ km s$^{-1}$, switching off the transverse components, $\delta v_{\perp,1} = \delta v_{\perp,2} =0$, where $\langle \cdots \rangle$ stands for the root-mean-squared average over time.
Figure \ref{fig:onlylongitude} presents mass-loss rate,
\begin{equation}
  \dot{M}_w = 4\pi\rho f v_{r} r^2,
  \label{eq:MLR}
\end{equation}
after time $t=3500$ minutes from the start of the simulation, where $f$ is the filling factor of open magnetic flux regions.
One can recognize negative $\dot{M}_w$ (dashed line) in the most part of the outer region, which indicates that the gas does not stream out but falls down to the surface; the acoustic wave cannot drive the solar wind only by itself.  

\begin{figure}[h]
\begin{center}
 \includegraphics[width=0.5\textwidth]{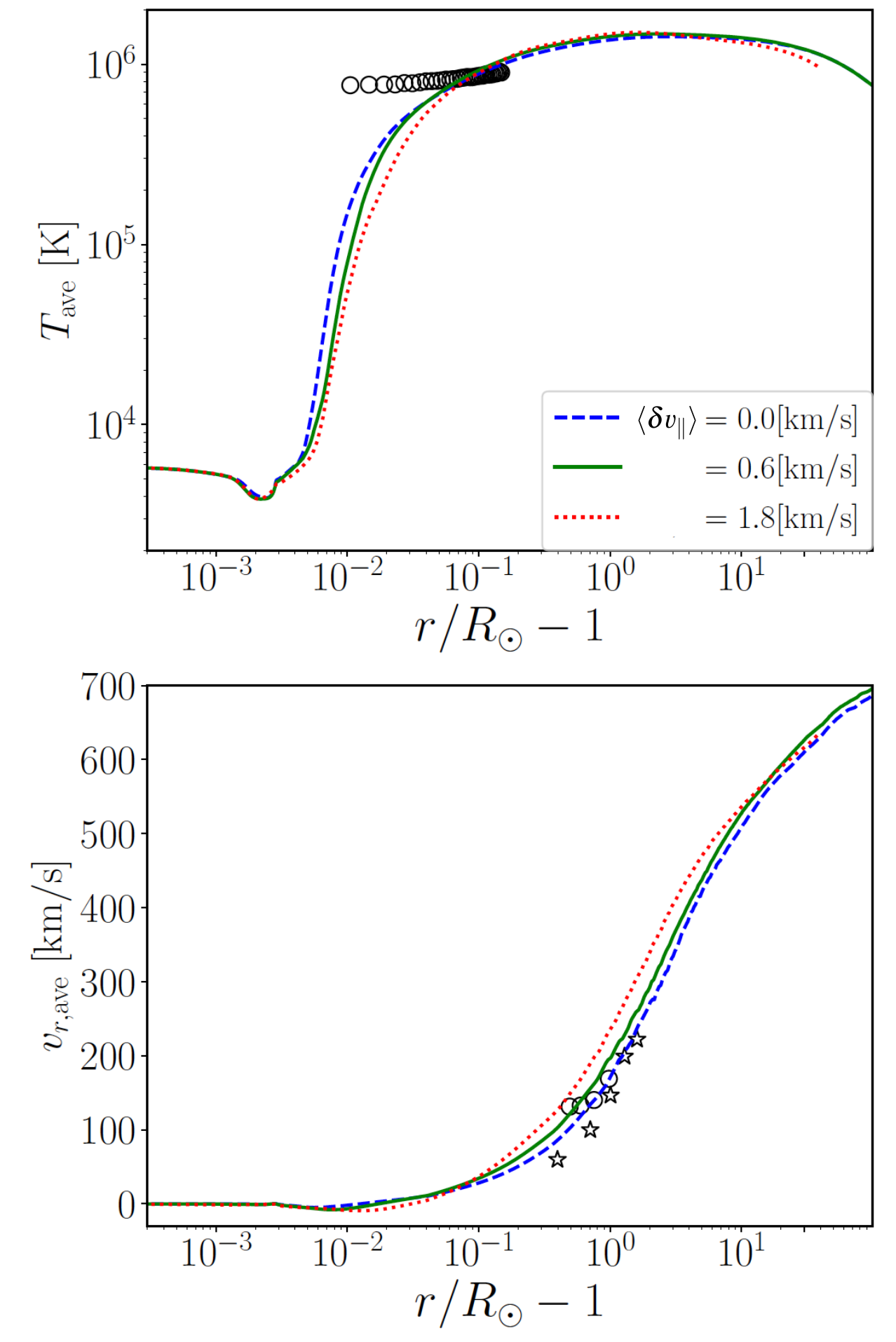} 
 \caption{Temperature (top) and radial velocity (bottom) of the cases with the longitudinal amplitude of $\langle \delta v_{\parallel}\rangle = 0$ km s$^{-1}$ (blue dashed), 0.6 km s$^{-1}$ (green solid) and 1.8 km s$^{-1}$ (red dotted) for the fixed transverse amplitude, $\langle \delta v_{\perp}\rangle= 0.6$ km s$^{-1}$. The circles in the top panel represent electron temperature observed by CDS/SOHO \citep{Fludra1999SSRv}. The stars \citep{Zangrilli2002ApJ} and the circles \citep{Teriaca2003ApJ} in the bottom panel  indicate proton outflow velocities observed in polar regions by SOHO.} \label{fig:T&v}
\end{center}
\end{figure}

\begin{figure}[h]
\begin{center}
 \includegraphics[width=0.6\textwidth]{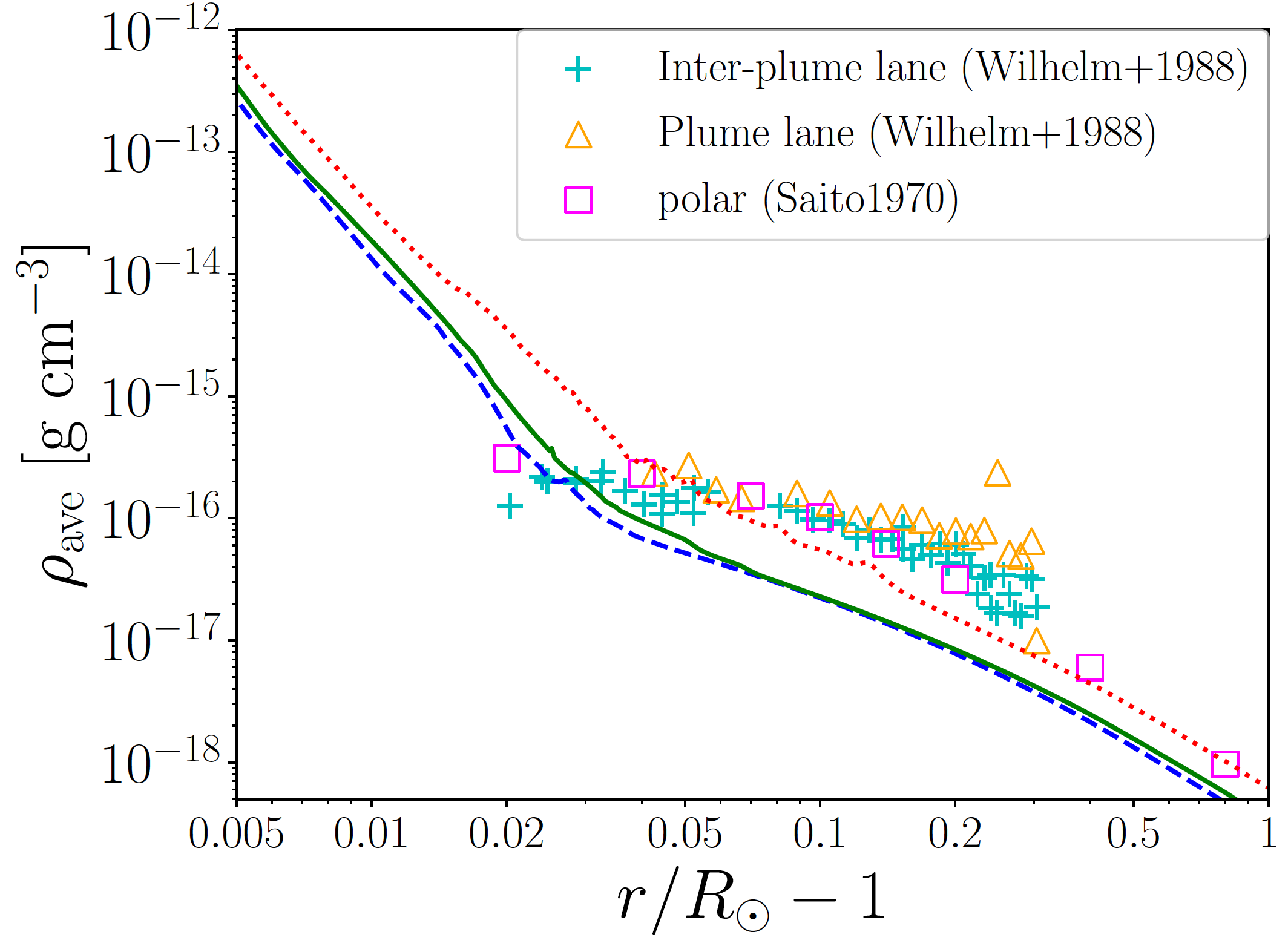} 
 \caption{Density of the same three cases as presented in Figure \ref{fig:T&v}. The linetypes are also the same as those in Figure \ref{fig:T&v}. The crosses and triangles are observed density \citep{Wilhelm1998ApJ} in interplumes and plumes, respectively. The squares show the density averaged over multiple solar eclipses during solar minimum phases \citep{Saito1970AnTok}.} \label{fig:rho}
\end{center}
\end{figure}

\begin{figure}[h]
\begin{center}
  \includegraphics[width=0.6\textwidth]{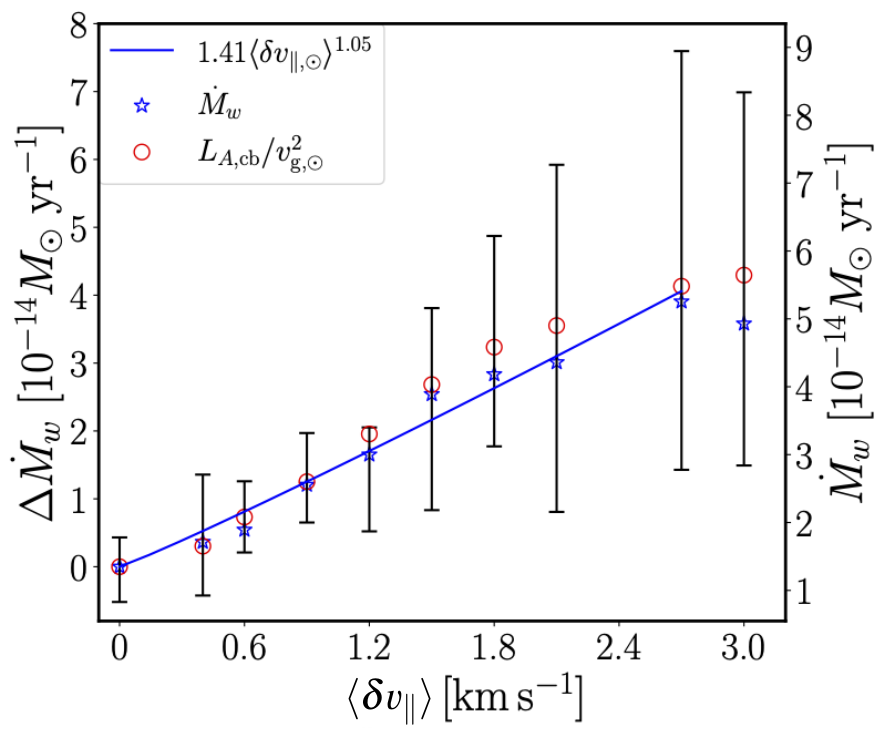}
  \caption{Mass-loss rate against input longitudinal-wave amplitudes on the photosphere.
   The right axis shows $\dot{M}_w$ (equation \ref{eq:MLR}). The left axis indicates the excess, $\Delta \dot{M}_w$, of  $\dot{M}_w$ from the value obtained for $\langle\delta v_{\parallel}\rangle = 0$.
   The blue star symbols are the values obtained from our numerical simulations.
   The red circles present the theoretical relation introduced by \citet[][see text]{Cranmer2011ApJ}. The blue solid line is the power-law fitting relation to $\Delta \dot{M}_w$ for $\langle\delta v_{\parallel}\rangle\le 2.7$ km s$^{-1}$.} \label{fig:MLR}
\end{center}
\end{figure}

Next, we study the effect of longitudinal perturbations in the presence of Alfv\'en waves. 
We vary the amplitude of the longitudinal perturbation at the photosphere in a wide range of $0$ km s$^{-1}$ $\le \langle \delta v_{\parallel}\rangle \le 3.0$ km s$^{-1}$ for a fix amplitude of the transverse components, $\langle \delta v_{\perp,1}\rangle = \langle \delta v_{\perp,2}\rangle = 0.6$ km s$^{-1}$.
Figure \ref{fig:T&v} compares the temperatures, $T$, (top) and radial velocities, $v_r$, (bottom) of three cases with different $\langle \delta v_{\parallel}\rangle = 0$ (blue dashed), $0.6$ (green solid), and $1.8$ km s$^{-1}$ (red dotted).
These three cases show quite similar profiles of $T$ and $v_r$.
In contrast, Figure \ref{fig:rho} shows that the density in the coronal region is higher for larger $\langle \delta v_{\parallel}\rangle$; the case with $\langle \delta v_{\parallel}\rangle =1.8$ km s$^{-1}$ yields $\approx 4$ times denser corona than the case with $\langle \delta v_{\parallel}\rangle =0$ km s$^{-1}$.

The difference in the coronal density directly leads to the change of the mass-loss rate of these cases.
Figure \ref{fig:MLR} shows that $\dot{M}_w$ obtained from the numerical simulations (blue stars) increases with $\langle\delta v_{\parallel}\rangle$ except for the range of $\langle\delta v_{\parallel}\rangle \ge 2.7$ km s$^{-1}$.
In addition, these numerical data are roughly reproduced by an analytic relation of $\dot{M_w}\approx L_{\rm A,cb}/v_{\rm g,\odot}^2$ \citep[red circles; ][]{Cranmer2011ApJ}, where $L_{\rm A,cb}$ is the Alfv\'enic Poynting flux at the coronal base \citep[see][for the specific expression]{Shimizu2022ApJ} and $v_{\rm g,\odot} = \sqrt{2GM_{\odot}/R_{\odot}}$ is the escape velocity. 
Since $v_{\rm esc,\odot}$ is constant in our setup, the dependence of $L_{\rm A,cb}$ on $\langle\delta v_{\parallel}\rangle$ indicates that the Alfv\'enic Poynting flux that reaches the corona increases by adding the photospheric longitudinal fluctuation even though the injected Alfv\'enic Poynting flux at the photosphere is the same.
While it is hard to understand this dependence because the longitudinal fluctuation, which generates acoustic-mode waves, does not directly contribute to Alfv\'enic Poynting flux, the connection between $\langle \delta v_{\parallel}\rangle$ and $L_{\rm A,cb}$ implies that something is happening between longitudinal and transverse waves in the chromosphere.

\begin{wrapfigure}{r}{7cm}
\begin{center}
\includegraphics[width=6cm]{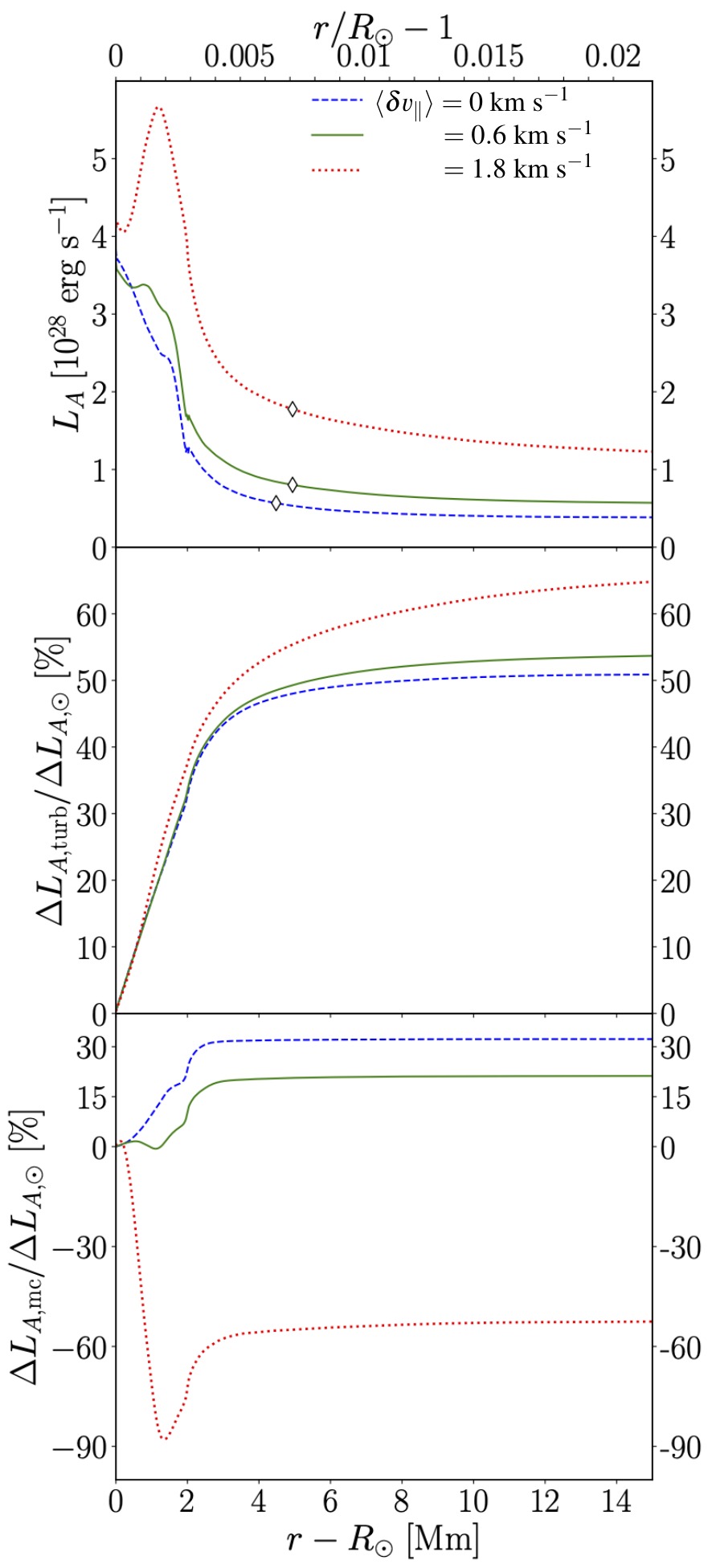} 
\caption{Comparison of the Alfv\'enic Poynting flux (top) and the relative contributions to the energy loss by the turbulent cascade (middle) and the mode conversion (bottom) of the three cases with different $\langle \delta v_{\parallel}\rangle$, where the linetypes are the same as in Figure \ref{fig:T&v}.} \label{fig:AlfvenPoy}
\end{center}
\end{wrapfigure}

To inspect the propagation and interaction of different modes of waves, we show the variation and dissipation of Alfv\'enic Poynting flux, $L_{\rm A}$, of the three cases with $\langle \delta v_{\parallel}\rangle= 0$, 0.6, and 1.8 km s$^{-1}$  from the photosphere to the low corona in Figure \ref{fig:AlfvenPoy}.
The top panel compares the vertical profiles of $L_{\rm A}$ of the three cases.
$L_{\rm A}$ of the two cases with zero and small $\langle \delta v_{\parallel}\rangle$ monotonically decreases by the dissipation of Alfv\'en waves.
In contrast, $L_{\rm A}$ of the case with the large $\langle \delta v_{\parallel}\rangle$ increases with height near the photosphere. 

The middle and bottom panels of Figure \ref{fig:AlfvenPoy} respectively presents the contributions of the turbulent cascade, $\Delta L_{\rm A,turb}$, and the mode conversion to longitudinal waves, $\Delta L_{\rm A,mc}$, in the dissipation of Alfv\'en waves, where both $\Delta L_{\rm A,turb}$ and $\Delta L_{\rm A,mc}$ are normalized by the Alfv\'enic Poynting flux at the photosphere, $L_{\rm A,\odot}$.  
In the case of no longitudinal fluctuation (blue dashed) both turbulent cascade and mode conversion play a significant role in the decrease of Alfv\'enic Poynting flux.
When the longitudinal perturbation with $\langle \delta v_{\parallel}\rangle = 0.6$ km s$^{-1}$ is input from the photosphere (green solid), the relative contribution of the mode conversion is reduced and the turbulent cascade is more important in the dissipation of Alfv\'en waves.
The mode conversion from transverse to longitudinal waves is suppressed because acoustic waves are pre-existing in the chromosphere by the longitudinal injection from the photosphere.
The tendency of the suppressed mode conversion is quite drastic in the case with large $\langle \delta v_{\parallel}\rangle = 1.8$ km s$^{-1}$ (red dotted), leading to a negative value of $\Delta L_{\rm A,mc}$, which indicates that the conversion from longitudinal waves to transverse waves is occurring in the chromosphere \citep{Schunker2006MNRAS,Cally2008SoPH}.
As a result, the Alfv\'enic Poynting flux increases there as shown in the top panel. 
Consequently, the Alfv\'en Poynting flux that reaches the corona is also larger for larger longitudinal fluctuation at the photosphere, giving larger mass-loss rate as discussed in Figure \ref{fig:MLR}.

\section*{Acknowledgment}
Numerical simulations in this work were partly carried out on Cray XC50 at Center for Computational Astrophysics, National Astronomical Observatory of Japan.
M.S. is supported by a Grant-in-Aid for Japan Society for the Promotion of Science (JSPS) Fellows and by the NINS program for cross-disciplinary study (grant Nos. 01321802 and 01311904) on Turbulence, Transport, and Heating Dynamics in Laboratory and Solar/ Astrophysical Plasmas: “SoLaBo-X.”
T.K.S. is supported in part by Grants-in-Aid for Scientific Research from the MEXT/JSPS of Japan, 17H01105, 21H00033, and 22H01263 and by Program for Promoting Research on the Supercomputer Fugaku by the RIKEN Center for Computational Science (Toward a unified view of the universe: from large-scale structures to planets, grant 20351188—PI J. Makino) from the MEXT of Japan.

\def\apj{{ApJ}}    
\def\nat{{Nature}}    
\def\jgr{{JGR}}    
\def\apjl{{ApJ Letters}}
\def\apjs{{ApJ Suppl.}}  
\def\aap{{A\&A}}   
\def\mnras{{MNRAS}}
\def\aj{{AJ}}
\def\ssr{{SSRv}}
\def\solphys{{Sol.Phys.}}
\let\mnrasl=\mnras



\end{document}